\documentclass[%
 reprint,
nofootinbib,
 amsmath,amssymb,
 aps,
 prx,
]{revtex4-2}

\usepackage{graphicx}
\usepackage{dcolumn}
\usepackage{bm}
\usepackage{hyperref}

\begin{document}

\preprint{imperial}

\title{Solving the Cosmic Coincidence Problem: The Locally Pumped Dark Energy Model}%

\author{Carlo~R.~Contaldi}

 \email{c.contaldi@imperial.ac.uk}
\affiliation{%
 Blackett Laboratory, Imperial College London,\\
Prince Consort Road, London SW7 2AZ, United Kingdom
}%

\author{Mauro Pieroni}

\email{
mauro.pieroni@csic.es
}
\affiliation{
Instituto de Estructura de la Materia (IEM), CSIC, Serrano 121, 28006 Madrid, Spain}%

\date{\today}

\begin{abstract}
We propose the Locally Pumped Dark Energy (LPDE) mechanism in which cosmic acceleration is triggered by the emergence of non-linear dark matter structure. In an effective-field-theory description, coarse-graining over the density contrast profile, whose short-wavelength modes grow during halo formation, induces a shift in the local equilibrium point of a second, sufficiently heavy scalar field $\chi$. At early times, the pump mechanism is negligible and $\chi$ remains fixed at the origin, contributing no DE. As structures form, the equilibrium value of $\chi$ is locally displaced within halos, generating a vacuum energy whose global contribution, in a mean-field picture, is controlled by the halo volume filling factor. If the $\chi$ field is sufficiently heavy, with a Compton wavelength limited by halo scales, its response is localised, and spatial gradients are exponentially suppressed on large scales. After volume-averaging over the halo population, the resulting contribution on large scales behaves as a homogeneous DE component. Using the halo mass function of a fiducial $\Lambda$CDM cosmology, we show that vacuum-energy domination generically emerges at $z\sim\mathcal{O}(1)$, naturally correlating cosmic acceleration with structure formation. For reference, we present an explicit realisation of such a mechanism and show that, by naturally featuring a transient acceleration epoch, it can be in excellent agreement with the most recent cosmological data, including the Dark Energy Spectroscopic Instrument (DESI).  
\end{abstract}

\maketitle

\section{Introduction}
\label{sec:intro}

The discovery that the expansion of the Universe is accelerating is one of the central puzzles of modern cosmology, first established by observations of distant type-Ia supernovae at the end of the 1990s~\cite{Riess:1998cb,Perlmutter:1998np}. Within the standard $\Lambda$CDM model, supported by precision measurements of the Cosmic Microwave Background (CMB), Baryon Acoustic Oscillations (BAO) and Large-Scale Structure (LSS)~\cite{Planck:2018vyg}, this acceleration is attributed to a cosmological constant $\Lambda$, whose observed value is many orders of magnitude smaller than typical vacuum-energy scales expected from quantum field theory. This gives rise to the well-known fine-tuning problem associated with the cosmological constant~\cite{Weinberg:1988cp}, as well as the coincidence problem: why does the onset of acceleration occur precisely at the epoch when the matter density has fallen to a value comparable to $\Lambda$ (see, e.g.,~\cite{Sahni:1999gb,Peebles:2002gy} for reviews)?

The idea that inhomogeneities might influence the large-scale expansion has been explored in the context of backreaction and averaging formalisms, most prominently in Buchert’s approach to averaging the Einstein equations over inhomogeneous domains~\cite{Buchert:1999er,Buchert:2001sa,Rasanen:2011ki}. In such models, additional terms arising from the variance of the local expansion and shear can, under certain conditions, mimic an effective Dark Energy (DE) component. However, there is ongoing debate over the quantitative importance of these effects in realistic cosmologies. A number of analyses have argued that backreaction remains perturbatively small in universes close to $\Lambda$CDM and cannot account for the observed acceleration (see, e.g.,~\cite{Ishibashi:2005sj,Green:2011qn,Green:2014aga,Clarkson:2011zq}). Nevertheless, this body of work highlights a conceptual link between non-linear structure formation and effective modifications of the large-scale expansion.

An intriguing empirical observation is that the onset of acceleration coincides with the cosmic epoch when most matter becomes locked into non-linear structures, such as galaxies, groups, and clusters. During matter domination, density perturbations grow, reach $\delta \sim \mathcal{O}(1)$ and eventually collapse into virialised halos. By redshift $z\sim 1$, a substantial fraction of the total matter density resides in collapsed objects. This suggests a possible direct connection between the emergence of non-linear structures and the onset of cosmic acceleration. If the late-time acceleration phase were triggered not by the homogeneous matter density $\rho_m(z)$, but instead by the clumped fraction of matter---encoded, for example, in the collapsed mass fraction $f_{\rm coll}(z)$---the apparent coincidence of these two phenomena would be significantly alleviated. Along similar lines, the authors in~\cite{Farrah:2023opk} have pointed to the apparent correlation of supermassive black hole mass function growth with the background accelerated expansion as evidence of coupling between small-scale inhomogeneities and DE.

In parallel, a rich literature has investigated Interacting DE (IDE) models, axion--quintessence couplings, and dark sector phase transitions. Coupled DE models typically introduce an interaction term proportional to the homogeneous Dark Matter (DM) density or to the DE field itself, yielding tracking solutions or modified growth of structures (see, e.g.,~\cite{Amendola:1999er,Boehmer:2008av,Valiviita:2009nu}, or~\cite{Khoury:2025txd} for a very recent proposal). Other approaches attempt to unify DM and DE within a single scalar field with multiple dynamical regimes or multi-branched potentials, including axion-like or monodromic models (e.g.,~\cite{Kobayashi:2018xvr,Marsh:2015xka}) or rely on scalar-tensor theories such as chameleon models~\cite{Khoury:2003aq,Brax:2004qh,Pietroni:2005pv}. Finally, a separate line of work examines inhomogeneous or clustering DE, analysing how DE fluctuations behave inside collapsing structures and how this affects halo formation and number counts~\cite{Nunes:2004wn,Mota:2004pa,Manera:2005ct,Baldi:2010vv}. 

Despite this variety of ideas, existing models generally rely on background quantities such as $\rho_m(z)$, on clustering of the DE component itself, or on generic DM--DE couplings that do not distinguish between linear and non-linear DM modes. To our knowledge, little attention has been paid to scenarios in which a DE phase transition is explicitly triggered by the non-linear component of the DM distribution, as quantified by $f_{\rm coll}(z)$ or by the variance of small-scale fluctuations. 

In this work, we propose a mechanism in which the growth of non-linear DM structure triggers late-time cosmic acceleration. For a concrete realisation, we assume DM is modelled as an axion-like scalar field $\phi$, which, once it begins coherent oscillations, behaves as Cold DM (CDM) on large scales and forms halos in the standard way (see, e.g,.~\cite{Marsh:2015xka} for a review). DE is provided by a second scalar field $\chi$ that is dynamically inert in the early, homogeneous Universe. As structure formation proceeds, short-wavelength axion fluctuations associated with virialised halos grow and act as an effective ``pump'' that modifies the potential of $\chi$. While this interaction is negligible when the Universe is smooth, it becomes important once a substantial fraction of matter resides in non-linear structures. Because $\chi$ is initially stabilised, it responds adiabatically to this change, remaining fixed at the origin at early times, but settling into a displaced minimum at late times. The resulting vacuum energy drives cosmic acceleration. In this picture, the onset of acceleration is governed by the non-linear evolution of DM, driven by gravity, rather than by the homogeneous background density, naturally linking the emergence of DE to the epoch of halo formation.

This mechanism parallels non-linear phonon pumping in driven condensed-matter systems, where a coherently excited fast mode modifies the effective potential of a slower structural mode, inducing a phase transition or non-linear phenomena~\cite{Forst:2011fk,Mankowsky:2014aya,Subedi:2014pua,Cavalleri:2018wmo}. In such systems, strong driving of an infrared-active phonon can reshape the lattice free energy and stabilise phases inaccessible in equilibrium. In the present context, small-scale axion fluctuations play the role of the pump mode, while $\chi$ acts as the structural mode whose vacuum configuration responds to the growth of non-linear DM clustering.

The manuscript is structured as follows. In section~\ref{sec:model}, we illustrate the mechanism, introduce the field setup, and motivate the phenomenological model using simple principles of EFTs resulting from coarse-graining over short modes. In section~\ref{subsec:chi_green_halos}, we describe how the local sourcing of vacuum energy can mimic a homogeneous, dynamical DE component. In section~\ref{sec:dynamics}, we show how the time-dependence of the halo mass function determines the dynamics of the model. We evaluate the background evolution and compare our results with recent observations. We summarise our results in section~\ref{sec:discussion} and discuss how this mechanism can be further developed and tested using existing and future observations.

\section{Locally Pumped Dark Energy (LPDE) mechanism}
\label{sec:model}

\subsection{Phenomenological model}

We consider two scalar fields evolving in a spatially flat FLRW spacetime with metric $g_{\mu\nu}$ and scale factor $a(t)$. The first field, $\phi$, is an axion-like scalar with mass $m_\phi$ that behaves as CDM once it begins coherent oscillations\footnote{Notice that the first field being an axion is not a strict requirement for the mechanism, which only requires a coupling between the second field and the local inhomogeneities that grow at late times.}. On large scales, its energy density redshifts as $a^{-3}$, while on small scales $\phi$ becomes non-linear and eventually collapses into virialised DM halos, with incoherent fluctuations dominating the field profile. The second scalar field, $\chi$, models the DE component only after significant non-linear structure has formed. At early times, $\chi$ is assumed to remain fixed near the origin, while at late times it tracks the minimum of an effective potential that is modified by the small-scale dynamics of the axion field.

The two scalar fields have canonical kinetic terms and are minimally coupled to gravity through an action of the form
\begin{equation}
S = \int \textrm{d}^4x \sqrt{-g}\,\left[
 \frac{M_{\rm Pl}^2}{2}R
 - \frac{1}{2}(\partial\phi)^2 
 - \frac{1}{2}(\partial\chi)^2 
 - V(\chi,\phi)
\right]\,,
\label{eq:action_full}
\end{equation}
where $M_{\rm Pl}$ is the reduced Planck mass and $R$ is the Ricci scalar. The potential $V(\chi,\phi)$ defines a complete theory with non-linear, bare terms that describe the interaction between $\chi$ and $\phi$. In our picture, the coarse-graining of non-linear, short-scale modes in the axion field induces an effective time dependence in the potential $V_{\rm eff} (\chi,\phi;a) $ such that in the homogeneous early Universe, the DE field has a simple quadratic potential,
\begin{equation}
V_0(\chi)=\frac12 m_\chi^2\chi^2\,,
\end{equation}
with $m_\chi^2>0$, so that $\chi$ is heavy, overdamped on cosmological timescales, and dynamically inert throughout radiation and early matter domination. In this regime, $\chi$ contributes no DE, and the background evolution reduces to the standard CDM. This does not avoid the fine-tuning problem, as we will discuss below. We assume the $\chi$-independent vacuum term is cancelled in the UV completion (set to zero by symmetry or sequestering); the LPDE contribution is additional and emerges only once non-linear structure develops. 

As structure formation proceeds, the axion field develops large short-wavelength fluctuations $\phi_S$ inside collapsed halos. In an effective-field-theory description, integrating out these non-linear modes induces a structure-dependent source term for the field $\chi$, shifting the location of the minimum of its effective potential. We assume that the parameters controlling the curvature of the potential are set by heavy physics and remain time independent, while all time dependence enters through the equilibrium position of $\chi$ (see section~\ref{subsec:classical_eft_shifted_quad} below). 

Phenomenologically, this effect can be modelled by the effective potential
\begin{equation}
\label{eq:Veff_quad}
V_{\rm eff}(\chi;a)
=
\frac12 m_\chi^2\Big[\chi-\chi_{\rm eq}(a)\Big]^2
+\rho_\chi(a),
\end{equation}
where effective mass $m_\chi$ now contains renormalisations induced by integrating out heavy fields and the shift $\chi_{\rm eq}(a)$ and uplift $\rho_\chi(a)$ encode the time dependence induced by coarse-grained short-scale axion correlators. 

\begin{figure}[t]
  \centering
  \includegraphics[trim=1.5cm 1.5cm 0.5cm 0cm,
    clip,width=0.45\textwidth]{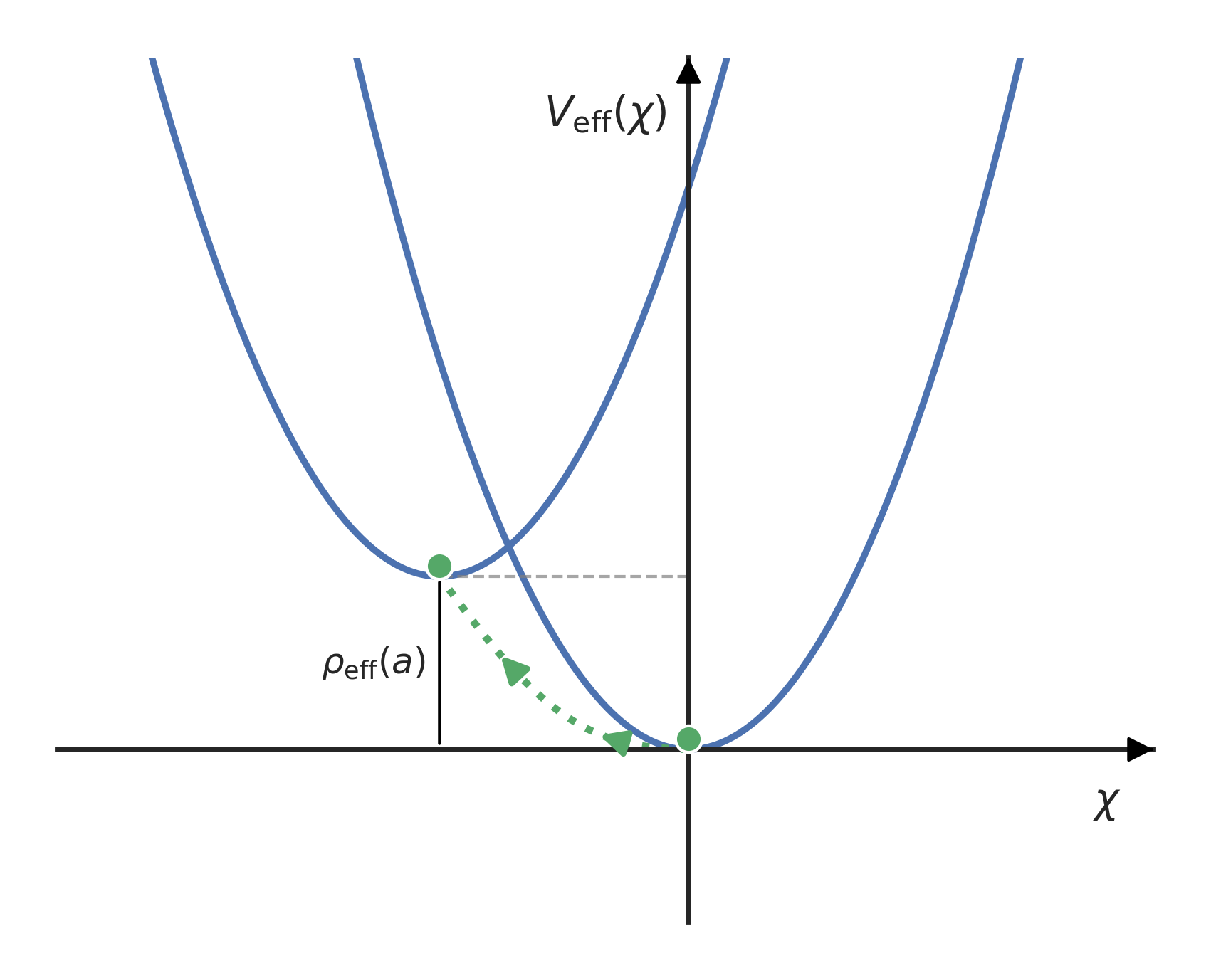}
  \caption{
  Illustration of the pump mechanism.
  The local equilibrium position shifts as non-linear structures form. The DE field $\chi$ tracks the evolving minimum adiabatically.
  }
  \label{fig:shifted_potential}
\end{figure}

As shown below, when the Universe is homogeneous and non-linear structures are absent, $\chi_{\rm eq}(a)\rightarrow 0$ and $\rho_\chi(a)\rightarrow 0$, so the field $\chi$ contributes no early DE. This automatically satisfies stringent early-Universe bounds (e.g., Planck CMB constraints~\cite{Planck:2018vyg}, Big Bang Nucleosynthesis (BBN) light-element abundances~\cite{Cyburt:2015mya}) without any engineered tuning of $V(\chi)$, ensuring a standard radiation- and matter-dominated background at early times. At late times, inside virialised halos, the DE field is pumped and moves adiabatically to an environment-dependent shifted equilibrium point, generating a local DE contribution to the density, see Fig.~\ref{fig:shifted_potential}. In section~\ref{subsec:chi_green_halos}, we argue why this local pumping mechanism can mimic a homogeneous, evolving DE component.

Unlike a fundamental cosmological constant, whose magnitude must be finely tuned to $\mathcal{O}(10^{-120})M_{\rm Pl}^4$, the vacuum energy in this framework is generated dynamically only after structure formation becomes significant, solving the coincidence problem. 

\subsection{EFT motivation for the locally pumped potential}
\label{subsec:classical_eft_shifted_quad}

The effective potential of Eq.~\eqref{eq:Veff_quad} can be understood as the leading long-wavelength description obtained after coarse-graining over short-distance axion fluctuations. The central EFT point is that, once short modes are integrated out, the long-wavelength theory for the heavy field $\chi$ contains \emph{all} local operators consistent with the symmetries, organised in an expansion in the ratio of long to short scales. The corresponding Wilson coefficients encode the net effect of matching to the UV completion and of coarse-grained short-mode dynamics. In particular, the absolute vacuum energy is not predicted within the EFT and must be fixed by matching; in what follows, we assume that the $\chi$-independent vacuum term vanishes in the UV completion, $\Lambda_0=0$, e.g., by a symmetry.

We decompose the axion-like DM field into long- and short-wavelength components,
\begin{equation}\label{eq:scale_split}
\phi(\mathbf{x},t)=\phi_L(t)+\phi_S(\mathbf{x},t),
\qquad
\langle \phi_S\rangle_S=0,
\end{equation}
where the long-wavelength mode is taken to be homogeneous and $\langle\cdots\rangle_S$ denotes coarse-graining over short modes defined with respect to a comoving cutoff $k_\ast$ (so that the physical cutoff is $k_\ast/a$). The long mode $\phi_L$ contains fluctuations coherent on large scales, while $\phi_S$ captures short-wavelength fluctuations that become non-linear and virialised inside collapsed structures. Such a long--short split underlies EFT descriptions of inflation and LSS, in which the influence of short-scale dynamics on long-wavelength observables is encoded through renormalised couplings and coarse-grained composite operators~\cite{Cheung:2007st,Baumann:2011su,Carrasco:2012cv,Burgess:2007pt,Senatore:2014via}.

For an axion, the underlying shift symmetry implies that interactions are periodic in $\phi/f_a$. Consequently, it is natural to build the short-mode ``pump'' from periodic operators such as $\langle 1-\cos(\phi_S/f_a)\rangle_S$ (or equivalently from short-mode energy-density operators). In the small-angle regime, this reduces to $\langle\phi_S^2\rangle_S/(2f_a^2)$, which we use below as a proxy for the coarse-grained short-scale amplitude. More generally, we denote by $\mathcal{O}_S(\phi_S)$ a renormalised, composite operator constructed from short modes (e.g., $\mathcal{O}_S\simeq\langle\phi_S^2\rangle_S$ or a linear combination including $\langle(\nabla\phi_S)^2\rangle_S$). After coarse-graining, the only information about the short modes that enter the long-wavelength background evolution is the corresponding local, scalar expectation value, which is a functional of time, 
\begin{equation}
\label{eq:OS_def}
\langle \mathcal{O}_S\rangle_S \equiv \mathcal{J}(a),
\end{equation}
whose time dependence arises from the growth of non-linear structures.

At energies and wavelengths relevant to the homogeneous background, $\chi$ is heavy and remains close to the minimum of its effective potential. The leading local operators involving $\chi$ and the pump source $\mathcal{J}(a)$ are therefore those of lowest mass dimension. Truncating at this order, the most general effective potential consistent with the assumed symmetries can be written as
\begin{equation}
\label{eq:Veff_EFT_truncated}
V_{\rm eff}(\chi;a)
=
\frac12 m_\chi^2\chi^2
+c_1\mu\mathcal{J}(a)\,\chi
+\frac12 c_2\,\mathcal{J}^2(a)+\Lambda_0\,,
\end{equation}
where $m_\chi^2>0$ is the effective mass of the $\chi$ field, $c_1$ and $c_2$ are (dimensionless) Wilson coefficients multiplying the leading purely short-mode operators, and $\mu$ is a mass scale that is tied to the non-linear coupling terms in the microphysical action. The EFT encapsulates the coupling between the $\chi$ field and the axion pumping mode once the ``fast'' axion oscillations, with frequency $m_\phi \gg m_\chi$ are averaged out in the coarse-graining. Higher-order operators, such as $\chi\,\mathcal{J}^2$ or $\mathcal{J}^3$, correspond to higher-dimension terms in the EFT and are suppressed by additional powers of the short-distance scale (set by the coarse-graining cutoff and/or the masses of heavy fields). We therefore retain only the terms in Eq.~\eqref{eq:Veff_EFT_truncated} as the minimal EFT parametrisation of the leading backreaction of non-linear axion fluctuations on the $\chi$ sector. Furthermore, we assume the bare cosmological constant that appears in the EFT is set to $\Lambda_0=0$ due to symmetry arguments. 

Minimising Eq.~\eqref{eq:Veff_EFT_truncated} gives the instantaneous local equilibrium position,
\begin{equation}
\label{eq:chi_eq_EFT}
\chi_{\rm eq}(a)=-\frac{c_1\mu}{m_\chi^2}\,\mathcal{J}(a),
\end{equation}
and completing the square yields the effective form in Eq.~\eqref{eq:Veff_quad}
with the vacuum energy at the minimum given by
\begin{equation}
\label{eq:rhoeff_EFT}
\rho_\chi(a)
=
\frac12\Big[c_2-\frac{c_1^2\mu^2}{m_\chi^2}\Big]\mathcal{J}^2(a).
\end{equation}
Eq.~\eqref{eq:Veff_quad} exhibits the desired EFT structure: all time dependence is
encoded in the displaced equilibrium position $\chi_{\rm eq}(a)$ (and hence in $\rho_\chi(a)$)
through the pump source $\mathcal{J}(a)$, while the curvature $m_\chi^2$ is set by heavy physics
and remains time independent. Provided the net coefficient in Eq.~\eqref{eq:rhoeff_EFT} is positive, which for a minimal assumption of $c_1 \lesssim 1$, $c_2 \lesssim 1$, implies that $\mu< m_\chi$, the structure-dependent contribution to the vacuum energy is positive and the model is stable.

Finally, we stress that this framework does not correspond to a scalar--tensor modification of gravity or to a conformal coupling of the form $g_{\mu\nu}\to A^2(\chi)g_{\mu\nu}$, nor to a direct coupling of $\chi$ to the trace of the (dark) matter stress-energy tensor. Such couplings underlie chameleon and symmetron scenarios~\cite{Hinterbichler:2010es,Hinterbichler:2011ca}, in which the scalar responds directly to the local density and generically mediates fifth forces that must be screened. By contrast, in the present framework, the activation of DE is controlled by coarse-grained correlators of non-linear DM fluctuations (encoded in $\mathcal{J}(a)$), rather than by the local density itself. Moreover, the field $\chi$ is assumed to remain heavy, $m_\chi\gg H$, so spatial variations are suppressed on cosmological scales and no long-range fifth-force phenomenology arises.

The presence of the Wilson coefficients and the $\Lambda_0$ term in the effective potential Eq.~\eqref{eq:Veff_EFT_truncated} emphasises that the cosmological-constant fine-tuning problem remains: within EFT, the coefficients of vacuum operators are UV sensitive and must be fixed by matching. In what follows, we treat $\rho_\chi(a)$ as the structure-generated vacuum component and choose parameters such that $\rho_\chi(a)\to 0$ in the homogeneous early Universe and becomes positive only once non-linear structure formation is significant.

\subsection{Initial conditions and mass hierarchy}
\label{subsec:initial_conditions}

In our framework, the distinct cosmological roles of $\phi$ and $\chi$ are fixed by their initial conditions and by a hierarchy of timescales.

The axion-like field $\phi$ follows the standard misalignment mechanism. During inflation, it is effectively light, $m_\phi\ll H_{\rm inf}$, and takes a nearly homogeneous value $\phi_{\rm ini}\neq0$ in our Hubble patch. Once $H(t_{\rm osc})\sim m_\phi$, the field begins coherent oscillations in its periodic potential and redshifts as $a^{-3}$, behaving as CDM on large scales. The mechanism requires that $\phi$ reproduces essentially standard non-linear structure formation on the halo scales relevant for the pump. For ultralight (``fuzzy'') masses, Lyman-$\alpha$ forest constraints typically require $m_\phi \gtrsim (2$--$5)\times10^{-21}\,\mathrm{eV}$ in canonical models (see, e.g.,~\cite{Irsic:2017yje,Armengaud:2017nkf,Rogers:2020ltq,Nori:2018pka}); for larger $m_\phi$ the halo mass function approaches that of CDM on the scales of interest. 

In contrast, the DE field $\chi$ is assumed to be stabilised at the origin at early times and does not undergo coherent oscillations when it becomes heavier than $H$. We therefore take $\chi_{\rm ini}\simeq0$ and $\dot\chi_{\rm ini}\simeq0$, so that $\chi$ is negligible through radiation domination and early matter domination\footnote{Notice that this is not a strict requirement for the mechanism to work. If the $\chi$ field undergoes coherent oscillations at sufficiently early times, it could, in principle, leave signatures in the CMB. However, as long as the energy associated with the oscillations is much smaller than the energy induced by the pump mechanism and than the axion energy density, the phenomenology will be unaffected.}. At late times, $m_\chi$ lies below the microscopic axion frequency given by $m_\phi$; however, this does not invalidate the construction because $\chi$ is not sourced by the raw oscillating field $\phi$, but by a coarse-grained composite operator which averages over the fast timescales. The relevant comparison is therefore between $m_\chi$ and the characteristic timescale for the gravitational evolution of the coarse-grained source, which will be of order $H^{-1}$, rather than between $m_\chi$ and $m_\phi$. When the local pump activates, $\chi$ tracks the instantaneous minimum $\chi_{\rm eq}(a)$ provided the minimum evolves slowly compared to the coarse-grained source dynamical time, i.e., $\textrm{d}\ln \chi_{eq}/\textrm{d}t\ll m_\chi$, so that deviations $\delta\chi\equiv\chi-\chi_{\rm eq}$ are rapidly damped. Thus, the required mass hierarchy, at late times, is $m_\phi\gg m_\chi\gg H_0$.

\section{From localised pumping to mean-field dark energy}
\label{subsec:chi_green_halos}

We now estimate the large-scale behaviour of the field $\chi$ when its effective potential is locally displaced inside virialised DM halos. The goal is to demonstrate that, after coarse-graining over the halo population, the contribution of $\chi$ to the background energy density is well-approximated by a homogeneous DE component, rather than a clustering one. 

Expanding around the local, halo-driven, equilibrium point $\chi_{\rm eq}$, the equation of motion for the heavy field may be written schematically as
\begin{equation}
\ddot\chi + 3H\dot\chi - a^{-2}\nabla^2\chi
+ m_\chi^2\big[\chi-\chi_{\rm eq}(\mathbf r)\big] = 0\,,
\label{eq:chi_eom_shifted}
\end{equation}
where we have introduced an explicit spatial dependence of the local equilibrium point.

In the adiabatic regime, in which the timescale associated with variations of $\chi_{\rm eq}$ is long compared to $m_\chi^{-1}$, time derivatives are subdominant and the equation reduces to the quasi-static form 
\begin{equation}
\label{eq:chi_static}
\big(\nabla^2-a^2 m_{\chi}^2\big)\chi(\mathbf r)
=
-a^2 m_{\chi}^2\,\chi_{\rm eq}(\mathbf r).
\end{equation}
In Appendix~\ref{sec:Yukawa}, we derive the solution to this equation, which exhibits Yukawa-like exponential suppression on large scales. It is useful to consider the mass scale required to suppress inhomogeneities on galaxy-scale halos. We have $\hbar H_0 \simeq 1.5\times10^{-33}\,\mathrm{eV}$, so that the Compton length of the $\chi$ field is
\begin{equation}
\lambda_\chi \equiv m_{\chi}^{-1}
\simeq 4.3\times10^{3}\,\mathrm{Mpc}\,
\left(\frac{H_0}{m_{\chi}}\right).
\end{equation}
In the ultra-heavy limit, the Compton wavelength of the field is far below typical halo scales, and the response of $\chi$ is tightly localised within the interior of individual halos. However, for the purposes of the LPDE mechanism, such an extreme hierarchy is not required. It is sufficient that $\chi$ remain heavy compared to the Hubble rate and to the inverse scales relevant for large-scale clustering, so that its response is quasi-local and its fluctuations are suppressed on cosmological scales. 

In what follows, we therefore allow an intermediate heavy-field regime with
\begin{equation}
\label{eq:LPDE_mass}
10^{4}H_0 \lesssim m_\chi \lesssim 10^{6}H_0,
\end{equation}
corresponding to a Compton wavelength in the approximate range
\begin{equation}
\label{eq:LPDE_wavelength}
10\,{\rm kpc}\lesssim \lambda_\chi \lesssim 1\,{\rm Mpc}.
\end{equation}
In this regime, the field remains non-clustering on large scales and continues to track its local equilibrium value adiabatically, while still permitting the LPDE profile to be more spatially extended than the DM profile within halos. This separation of scales is useful phenomenologically, since it allows the pumped component to remain dynamically subdominant within virialised objects even when its volume-averaged cosmological contribution is significant. We defer the detailed halo-scale motivation for this intermediate mass range to Appendix~\ref{sec:DE_density_profile}.

More generally, on scales $L\gg \lambda_\chi$, spatial gradients are suppressed, and the coarse-grained field is well approximated by
\begin{equation}
\bar\chi \equiv \langle \chi(\mathbf r)\rangle \simeq \langle \chi_{\rm eq}(\mathbf r)\rangle,
\end{equation}
so that the cosmological background depends only on the mean halo abundance and the typical internal pumping amplitude $\mathcal{J}(a)$, rather than on the detailed spatial distribution of individual halos.

To make the volume-averaging explicit and understand how an effectively homogeneous DE component emerges, we consider a population of halos labelled by $i$, with physical radius $R_{h,i}$ and equilibrium shift $\chi_{\rm eq}({\mathbf r})$ inside each halo. In the adiabatic regime, the local energy density in the $\chi$ sector inside a halo is well approximated by its vacuum value,
\begin{equation}
\rho_\chi(\mathbf r)\simeq \rho_\chi\big(\chi_{\rm eq}(\mathbf r)\big),
\end{equation}
while outside halos $\chi_{\rm eq}\simeq 0$ and $\rho_\chi\simeq 0$. To calculate the cosmological contribution of $\rho_\chi$, we need to average the individual halo contributions over the density of halos in a cosmological volume.

We define the mean $\bar \rho_{\chi,h}$ energy density associated with a halo of mass $M$, radius $R_h(M,z)$, and volume\footnote{As discussed in Appendix~\ref{sec:DE_density_profile}, technically, the volume $V_{\chi}$, where the $\chi$ field contributes as DE, is larger than the halo volume $V_h$. For simplicity, in this section, we neglect this distinction.} $V_h=4\pi R_h^3/3$ with 
\begin{equation}\label{eq:mean_rho}
    \bar \rho_{\chi,h}(M,z)\equiv \frac{1}{V_h(M,z)}\int_{V_h} \textrm{d}^3 r\,\rho_\chi({\mathbf r}|M,z).
\end{equation}
Then the DE contribution from the halo is 
\begin{equation}
E_{\chi,h}(M,z) \equiv V_{h}(M,z)\,\bar \rho_{\chi,h}(M,z).
\label{eq:Echi_halo}
\end{equation}
Summing over halos $i$ in a large comoving volume $V$ and dividing by $V$ gives the coarse-grained mean energy density
\begin{eqnarray}
\label{eq:rhochi_halomf}
\bar\rho_\chi(z)
&\simeq&
\frac{1}{V}\sum_i E_{\chi,h_i}\,,\\
&\simeq
&\int \textrm{d}M\;\frac{\textrm{d}\tilde n}{\textrm{d}M}(M,z)\,V_h(M,z)\,\bar\rho_{\chi,h}(M,z),\nonumber
\end{eqnarray}
where $\textrm{d}\tilde n/\textrm{d}M$, the halo mass function, is the (physical) halo number density per unit mass. Eq.~\eqref{eq:rhochi_halomf} exhibits a key feature of this mechanism: although the pumping is localised inside halos, the cosmological background is sensitive only to the volume-averaged contribution, which depends on the halo abundance, the characteristic halo volumes, and the local coarse-grained pump term determined by Eq.~\eqref{eq:chi_eq_EFT}. 

In realistic models, the value of $\chi_{\rm eq}$ will exhibit significant dependence on the local environment and on the halo formation history. However, the fact that long-wavelength fluctuations are suppressed means that, on scales larger than a reference halo scale, $\bar \rho_\chi$ will always appear as a homogeneous, non-clustering, evolving DE component. Observational consequences will exist on sub-halo scales, and we expect the study of sub-halo structure dynamics to be a fruitful avenue for falsifying this family of models.

In the following, we are interested in tracking the halo number density within a mass range associated with objects that are virialised by a certain redshift $z$. To estimate this quantity, we can integrate the halo mass function $\textrm{d}\tilde n/\textrm{d}M$ across a mass range $[M_{\rm min},M_{\rm max}]$, covering, for example, dwarf galaxies to clusters \footnote{In principle, we could choose a smaller value for $M_{\rm min}$ to include ultra-faint galaxies, which can could be up to a few order of magnitudes lighter~\cite{Kirby:2013isa, Koposov:2015cua, DES:2015txk, DES:2015zwj}.}, $10^{7}M_\odot\to 10^{16}M_\odot$. These limiting masses are free parameters of the model and are related, phenomenologically, to the coarse-graining scale defining the split between long- and short-modes in Eq.~\eqref{eq:scale_split}.

For simplicity, to explore the viability of this mechanism and to connect with analytic intuition, we consider a minimal approximation in which the interior pumping amplitude is approximately independent of halo mass and scales only with time. This provides a proxy for how the pump amplitude can grow as collapsed structures become more concentrated over time without introducing a complete micro-dynamical model for halo formation.

Under these assumptions 
\begin{align}
  \bar\rho_\chi(z) &= \bar\rho_{\chi,h}(z)\, \int^{M_{\rm max}}_{M_{\rm min}} \textrm{d}M\;\frac{\textrm{d}\tilde n}{\textrm{d}M}(M,z)\,V_h(M,z),\nonumber\\
  &=\bar\rho_{\chi,h}(z)\, f_V(z),
\label{eq:rhochi}
\end{align}
where $f_V(z)$ is the physical volume filling factor in halos within the mass range of interest. This factor is related to the fraction of mass in collapsed objects within this mass range, defined as 
\begin{align}
f_{\rm coll}(z)
&\equiv \frac{\rho_m^{\rm coll}(z)}{\rho_m(z)},\\
&\equiv \frac{1}{\rho_{m}(0)}\int_{M_{\rm min}}^{M_{\rm max}} \textrm{d}M\; M\,\frac{\textrm{d}n}{\textrm{d}M}(M,z),
\label{eq:fcoll_hmf}
\end{align}
where now, $\textrm{d}n/\textrm{dM}$ is the comoving halo mass function\footnote{Notice that the integral in Eq.~\eqref{eq:fcoll_hmf} is now expressed in terms of the \emph{comoving} number density, so that the volume and $\rho_{m}$ scaling with $z$ cancel out.}. This quantity captures the growth of non-linear structures and is sensitive to the halo abundance and its redshift evolution.

To calculate $f_V(z)$, one can adopt the conventional overdensity threshold~\cite{Davis:1985rj, 
Lacey:1994su,
Tinker:2008ff,
j_binney_galactic_1987, 
Mo_van_den_Bosch_White_2010} 
\begin{equation}
M = 200\,\rho_c(z)\, V_h(M,z) ,
\label{eq:R200_def}
\end{equation}
meaning that halos require a mean matter overdensity $\bar\rho_{m,h}$, calculated as in Eq.~\eqref{eq:mean_rho}, of 200 compared to $\rho_c(z)$. We then have 
\begin{equation}
f_V(z)
\equiv
\int_{M_{\rm min}}^{M_{\rm max}} \textrm{d}M\;
\frac{\textrm{d}\tilde n}{\textrm{d}M}(M,z)\,
V_h(M,z),
\label{eq:fV_def_phys}
\end{equation}
and, using Eq.~\eqref{eq:R200_def} to eliminate the volume term, one can identify
\begin{align}
f_V(z)
&=
\frac{1}{200\,\rho_c(z)}\int_{M_{\rm min}}^{M_{\rm max}} \textrm{d}M\;
M\,\frac{\textrm{d}\tilde{n} }{\textrm{d}M}(M,z),
\nonumber\\
&=
\frac{\Omega_m(z)}{200}\,f_{\rm coll}(z),
\label{eq:fV_identity}
\end{align}
where $\Omega_m(z)=\rho_m(z)/\rho_c(z)$ is the total matter density parameter. Eq.~\eqref{eq:fV_identity} reinforces that the geometric suppression of the halo-localised pumping is set by the collapsed fraction and the ratio of the mean matter density to the halo interior density.

\subsection{Modelling the pump mechanism from non-linear structure formation}
\label{sec:ST_formalism_for_fV}

To compute the global, homogeneous DE density generated by halo-localised pumping, we require the abundance of collapsed objects as a function of halo mass and redshift. We therefore model the halo population using a standard halo mass function, calibrated to the linear matter power spectrum of a fiducial $w$CDM cosmology. In practice, the linear spectrum $P(k,z)$ may be obtained from a Boltzmann solver such as \texttt{CAMB}~\cite{Lewis:1999bs}, and used to construct the variance of the smoothed density field on a Lagrangian mass scale $M$. 

Defining a comoving top-hat smoothing radius $R(M)$ using
\begin{equation}
M = \frac{4\pi}{3}\,\rho_{m}(0)\,R^3,
\qquad
R(M)=\left(\frac{3M}{4\pi\rho_{m(0)}}\right)^{1/3},
\end{equation}
where $\rho_{m}(0)$ is the energy density of matter today at $z=0$, the linear variance is
\begin{equation}
\sigma^2(M,z)=\int_0^\infty \frac{\textrm{d}k}{2\pi^2}\,k^2\,P(k,z)\,W^2(kR),
\end{equation}
with the window function
\begin{equation}
W(x)=\frac{3(\sin x-x\cos x)}{x^3},
\label{eq:sigma_def}
\end{equation}
and the corresponding peak height is
\begin{equation}
\nu(M,z)\equiv \frac{\delta_c}{\sigma(M,z)},
\label{eq:nu_def}
\end{equation}
where $\delta_c\simeq 1.686$ is the linear threshold for spherical collapse.

We write the comoving halo mass function in the universal form~\cite{Press:1973iz}
\begin{equation}
\frac{\textrm{d}n}{\textrm{d}M}(M,z)
=
\frac{\rho_{m}(0)}{M}\,f(\nu)\,\left|\frac{\textrm{d}\nu}{\textrm{d}M}\right|,
\label{eq:hmf_universal}
\end{equation}
where $f(\nu)$ is the multiplicity function. For definiteness, one may adopt the Sheth-Tormen parametrisation~\cite{Sheth:1999mn},
\begin{equation}
f_{\rm ST}(\nu)=A\sqrt{\frac{2q}{\pi}}
\left[1+(q\nu^2)^{-p}\right]\nu\,\exp\!\left(-\frac{q\nu^2}{2}\right),
\label{eq:sheth_tormen}
\end{equation}
with canonical parameters $p\simeq 0.3$, $q\simeq 0.707$ and $A\simeq 0.322$.

\begin{figure}[t]
  \centering
  \includegraphics[width=\linewidth]{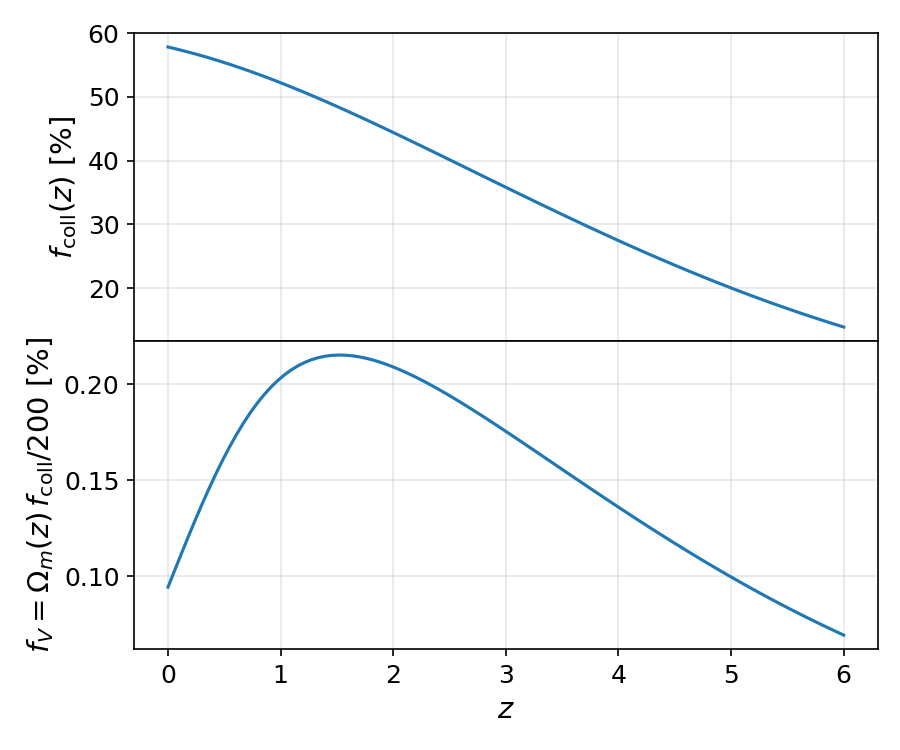}
  \caption{
  Collapsed fraction (top) and volume filling fraction (bottom) as a function of redshift for a representative PLANCK-like cosmology calculated using the Sheth-Tormen parametrisation for the halo mass function.
  \label{fig:f_coll}}
\end{figure}

In Fig.~\ref{fig:f_coll}, we show the evolution of $f_{\rm coll}$ and $f_V$ as a function of redshift for a representative PLANCK-like cosmology~\cite{Planck:2018vyg} (i.e., the background evolution is not modified by the LPDE, such a self-consistent analysis is postponed to Sec.~\ref{sec:dynamics}) using the formalism described above. Notice that while $f_{\rm coll}$ keeps growing with time and will eventually saturate, $f_V$ reaches a maximum and then decreases because of the late time dilution induced by the accelerated expansion, which sees $\Omega_m(z)$ decreasing at late times. 
 
In the simplified halo-localised pumping picture of Eq.~\eqref{eq:rhochi}, the global DE density is controlled by $f_V(z)$ and the time-dependence of the mean interior equilibrium shift $\chi_h(z)$ for halos. The time dependence of the DE component is therefore inherited from the evolving halo mass function and time-dependence of $\chi_h(z)$.

\subsection{Redshift dependence of $\rho_{\chi_h}$}

Physically, $\rho_{\chi,h}$ should be controlled by the degree of non-linear evolution within individual halos after they first reach the critical overdensity for collapse. As halos virialise and their inner structure becomes increasingly compact, gradients and/or local variance in the underlying field grow, enhancing the coarse-grained pumping amplitude $\mathcal{J}_h^2(z)$.

A first-principles calculation of this non-linear evolution, including its dependence on internal halo structure, would require detailed modelling of the fully non-linear dynamics. Rather than attempting such a calculation here, we adopt a simple phenomenological parametrisation that captures the expected late-time growth of the pumping mechanism. We assume a power-law dependence on the scale factor,
\begin{equation}
  \frac{\mathcal{J}_h(z)}{\mathcal{J}_h(0)} = \left[\frac{a(z)}{a(0)}\right]^{p_{\chi}},
\end{equation}
 where $p_{\chi}$ is a free parameter. For $p_{\chi}>0$, the pumping effect is suppressed at early times and grows towards low redshift, providing a proxy for the increasing compactness and non-linear development of halo structure. This parametrisation allows us to explore the impact of time-dependent pumping while keeping the model analytically simple and computationally tractable.

Since the induced vacuum energy scales quadratically with the coarse-grained amplitude, using Eq.~\eqref{eq:rhochi}, we get that the resulting DE density evolves as
\begin{equation}
\frac{\bar \rho_\chi(z)}{\bar\rho_{\chi,h}(0)}
= \frac{f_V(z)}{(1+z)^{2p_{\chi}}}\,.
\end{equation}
This provides a minimal phenomenological realisation of the LPDE mechanism.

\section{Background evolution and effective equation of state}
\label{sec:dynamics}

In the halo--localised LPDE mechanism, the scalar field $\chi$ is canonical and remains heavy, and therefore its kinetic energy remains negligible throughout the cosmologically relevant epochs. This means that while the local Equation of State (EoS) parameter satisfies $w_\chi \simeq -1$, the theory never enters a phantom regime and remains free of ghost instabilities.

As detailed above, a homogeneous DE density is obtained by volume averaging
over the halo population. Once this is done, on cosmological scales where the FLRW limit is assumed, the background history is governed by the usual Friedmann equation
\begin{equation}
H^2(z)
=\frac{8\pi G}{3}
\left[
\rho_r(z)+\rho_b(z)+\rho_m(z)+\bar\rho_\chi(z)
\right],
\label{eq:friedmann_background_new}
\end{equation}
where $\rho_r\propto a^{-4}$, $\rho_b\propto a^{-3}$, and $\rho_m\propto a^{-3}$ denotes the homogeneous DM (axion) density. Since $\bar\rho_\chi(z)$ arises from structure formation, it vanishes at early times and grows during halo collapse. This growth is not only driven by the background expansion, but also by the interactions between the DM and DE sectors, since the small-scale DM, driven by gravitational collapse, is injecting energy into the DE field $\chi$. The Bianchi identities of the complete theory still imply conservation equations for both fluids, which would be coupled via an interaction term $Q(a,k)$ as
\begin{align}
&\dot{\bar\rho}_\chi
+3H\bigl(1+w_{\chi}(z)\bigr)\bar\rho_\chi=Q(a,k),\\
&\dot{\rho}_m
+3H\rho_m=-Q(a,k).
\label{eq:rhodot}
\end{align}
Note that, because of the backreaction of non-linear matter structures, in the LPDE model, the forcing is strongly scale dependent, unlike conventional IDE parametrisations, where 
$Q$ is built from homogeneous densities (see, e.g.,~\cite{vanderWesthuizen:2025rip, vanderWesthuizen:2025vcb, vanderWesthuizen:2025mnw} for a comprehensive review). In the LPDE model, energy transfer is controlled by non-linear short-scale correlators and therefore inherits an intrinsically scale-dependent activation tied to halo formation.

In our EFT treatment, the system is not closed, and we model the growth of $\bar\rho_\chi$ phenomenologically through the halo mass function, volume effects, and non-linear growth. To link to observations, we define an effective EoS parameter using the conventional definition
\begin{equation}
\dot{\bar\rho}_\chi
+3H\bigl[1+w_{\rm eff}(z)\bigr]\bar\rho_\chi=0,
\label{eq:weff_def}
\end{equation}
which yields
\begin{equation}
w_{\rm eff}(z) = -1 -\frac{1}{3}\frac{\textrm{d}\ln\bar\rho_\chi}{\textrm{d}\ln a}.
\label{eq:weff_log}
\end{equation}
Although the underlying field $\chi$ has a canonical action and never behaves as a phantom field, the effective EoS parameter will generally display phantom behaviour when the structure formation is driving rapid growth of in $\bar\rho_\chi$. Conversely, if the volume-averaged DE contribution decreases, then the effective EoS parameter could be driven to $w_{\rm eff}>-1$. A general property of the LPDE model is therefore that $w_{\rm eff}$ is not constant. Moreover, because of gradient suppression, the effective DE fluid $\bar\rho_\chi$ does not cluster and remains smooth on large scales. This is despite its EoS parameter deviating from pure cosmological constant $w=-1$.

With these principles in mind, we now examine the background expansion implied by the LPDE prescription. To do this, we must solve for a self-consistent relation between the effective DE density and the halo volume-filling factor $f_V(z)$. For notational clarity, we identify $\bar\rho_\chi(z)\equiv\rho_{\rm DE}(z)$ below.

The key numerical difficulty is that $f_V(z)$ depends on the cosmology through the background expansion and the linear matter power spectrum $P(k,z)$, while $\rho_{\rm DE}(z)$ simultaneously feeds back on the background. We therefore adopt an iteration scheme to calculate self-consistent background solutions using convergence in the function $\ln\rho_{\rm DE}(a)$:
\begin{align}
\ln\rho_{\rm DE}^{(n+1)}(z_i)
&=(1-\epsilon)\,\ln\rho_{\rm DE}^{(n)}(z_i)
+\epsilon\,\ln\rho_{\rm DE}^{\rm raw}(z_i),
\end{align}
with a sample of redshifts $z_i\in\{z_{\rm list}\}$. The under-relaxation parameter $\epsilon$ is introduced to ensure smooth convergence. At each iteration $n$:
\begin{enumerate}
\item Given $\ln\rho_{\rm DE}^{(n)}(z_i)$ on a late-time redshift grid $z_{\rm list}$, we construct a smooth representation $\ln\rho_{\rm DE}^{(n)}(\ln a)$ using a smoothing spline in $\ln a$.
\item We convert this smooth density history into a tabulated EoS parameter using 
\begin{equation}
w_{\rm eff}^{(n)}(a) \equiv -1 - \frac{1}{3}\,\frac{\textrm{d}\ln\rho_{\rm DE}^{(n)}}{\textrm{d}\ln a}\,,
\end{equation}
where the derivative is obtained analytically from the spline.
\item We provide the resulting tabulated $w_{\rm eff}^{(n)}(a)$ directly to \texttt{CAMB} (using the PPF DE module) to obtain the background $H(z)$ and a linear matter power spectrum interpolator $P(k,z)$ on $z_{\rm list}$.
\item Compute $f_V(z)$ using the procedure described in Section~\ref{sec:ST_formalism_for_fV}, and get the the raw update $\rho_{\rm DE}^{\rm raw}(z)\propto f_V(z)/(1+z)^{2p}$.
\end{enumerate}
The iteration is terminated when the maximum change $\max_i|\Delta\ln\rho_{\rm DE}(z_i)|$ falls below a fixed (relative) tolerance of $10^{-3}$ after a minimum number of iterations (which we set to three).

\begin{figure*}[t]
  \centering
  \includegraphics[width=\textwidth]{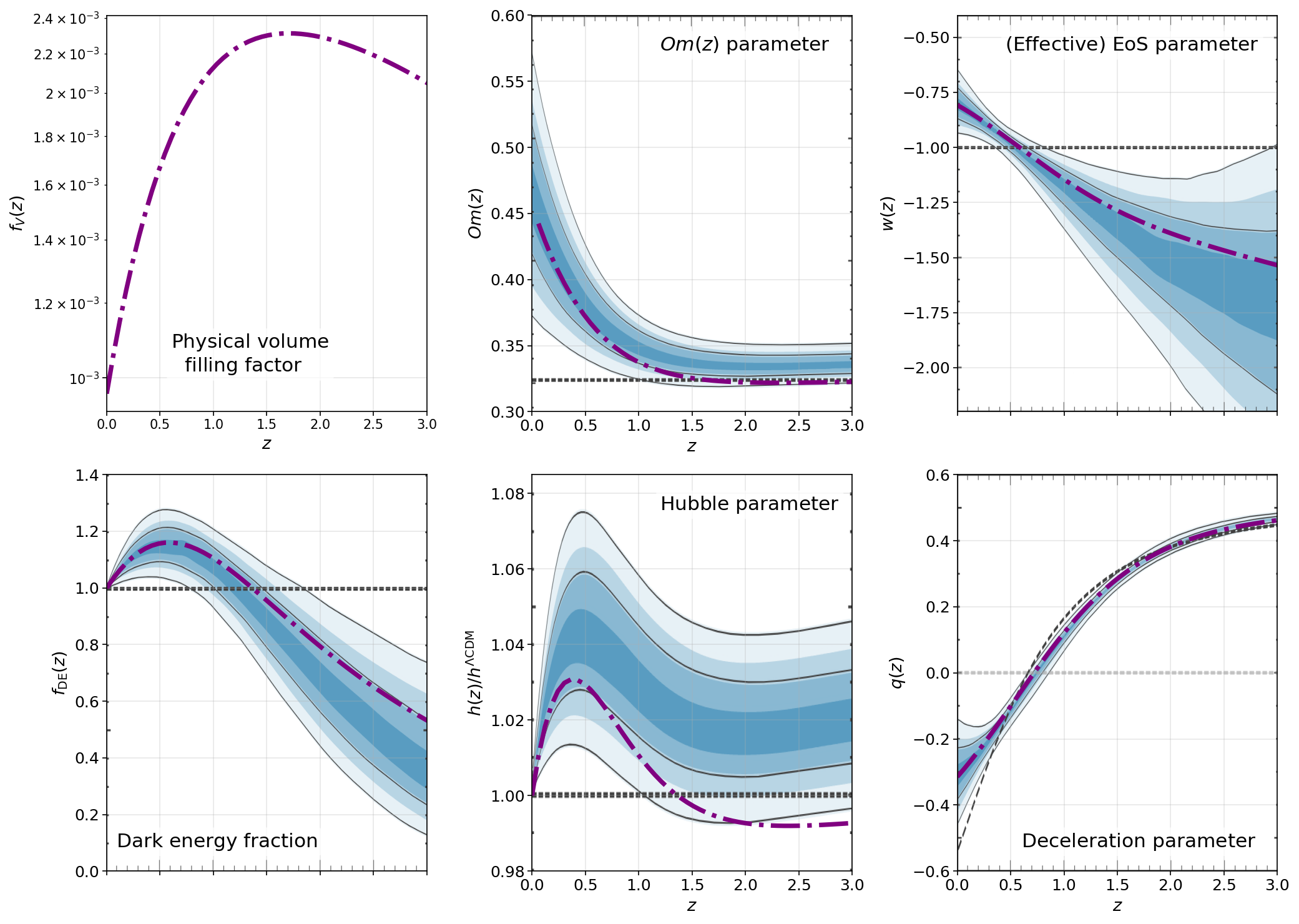}
  \caption{LPDE predictions (purple dash-dotted line) for $f_V(z)$ (top-left panel), $f_{\rm DE}(z) $ (bottom-left panel), $Om(z)$ (top-centre panel), the (normalised) Hubble parameter $h(z) \equiv H(z)/H_0$ normalized to the $\Lambda$CDM value (bottom-centre panel), the (effective) EoS parameter $w_{\rm eff}(z)$ (top-right panel), and the deceleration parameter $q(z)$ (bottom-right panel) as functions of redshift (see main text for the definitions of these quantities) for $p_{\chi} = 1/2$, $M_{\rm Min} = 10^{8} M_{\odot}$, $M_{\rm Min} = 10^{18}M_{\odot}$. In these plots, the colourful cyan bands represent the 1 to 3$\sigma$ (in progressively lighter shades of cyan) constraints from~\cite{DESI:2025fii} (see main text) and, for reference, dark grey dashed lines are used to highlight the $\Lambda$CDM predictions. }
  \label{fig:cosmo_summary}
\end{figure*}

Two practical stabilisers are employed. First, the tabulated $w_{\rm eff}(a)$ is padded to early times by setting $w_{\rm eff}(a)=-1$ for $a<a_{\rm anchor}$ (corresponding to $z>z_{\max}=6$ of the halo grid), since our halo-based prescription is only intended to describe the late-time regime. Second, during the first few iterations, we optionally clip extreme values of $w_{\rm eff}(a)$ to prevent transient numerical noise in the spline derivative from driving the Boltzmann integration into unphysical regions; this clipping is not required once the fixed point is reached. This procedure preserves the intended property that the LPDE component is treated as a smooth background (no DE perturbations are introduced), while still capturing the cosmology dependence of $f_V(z)$ through $H(z)$ and $P(k,z)$.

In Fig.~\ref{fig:cosmo_summary}, we show the behaviour of a set of physically relevant quantities. In all these plots, the results of the LPDE model with $M_{\rm min} = 10^{8} M_{\odot}$, $M_{\rm max} = 10^{18} M_{\odot}$, and $p_{\chi} = 1/2$, are plotted with dash-dotted purple lines (see Appendix~\ref{sec:parameter_dependence} for different parameter points and for a more details on the impact of the different parameters on the results shown in Fig.~\ref{fig:cosmo_summary}). In most panels, the LPDE are compared with the most recent Gaussian Process (GP) regression non-parametric constraints (1 to 3-$\sigma$ levels in progressively lighter shades of cyan) from~\cite{DESI:2025fii}, which combine the DESI DRII data with the Union3 data~\cite{Rubin:2023jdq} and CMB data from Planck~\cite{Planck:2018nkj} (see~\cite{DESI:2025fii} for more details), together with the $\Lambda$CDM predictions in dashed dark grey lines. More in detail, in the top-left panel, we show $f_V(z)$, in the bottom-left panel, we show the DE density normalised to its current value
\begin{equation}
\label{eq:fde}
f_{\rm{DE}}(z) \equiv \frac{\rho_{\rm{DE}}(z)}{\rho_{\rm{DE},0}} = \exp\left[ 3 \int_0^z  \frac{1 + w_{\rm eff}(z^\prime)}{1 + z^\prime} \, , \textrm{d}z^\prime\right] \, ,
\end{equation}
in the top-centre panel, we show the $Om(z)$ diagnostic~\cite{Sahni:2008xx}
\begin{equation}
    Om(z) \equiv \frac{h^2(z) - 1}{(1 + z)^3 - 1} \; , 
\end{equation}
where $h(z) \equiv H(z)/H_0$ is the (normalised) Hubble parameter, whose value, normalised to the $\Lambda$CDM value, is shown in the bottom-centre plot. In the top-right plot, we show the EoS parameter $w_{\rm eff}$, and in the bottom-right plot, we show the deceleration parameter
\begin{equation}
\label{eq:q}
q(z) \equiv \frac{\rho_{\rm{DE}}(z)}{\rho_{\rm{DE},0}} = \exp\left[ 3 \int_0^z  \frac{1 + w_{\rm eff}(z^\prime)}{1 + z^\prime} \, , \textrm{d}z^\prime\right] \, .
\end{equation}
For all parameters, with the only exception of $Om(z)$ and $h(z)/h^{\Lambda \rm CDM}$, where the early time predictions are slightly lower than the measured values\footnote{Notice that while $h(z)$ is independent of $H_0$, a relative change of $H_0$ between the $\Lambda$CDM model and the LPDE model will affect the purple line in the bottom-centre plot. In particular, we found that an order percent difference between the two models (with a slightly larger value for $\Lambda$CDM) is sufficient to reconcile theory with observations.}, we notice an excellent agreement between the LPDE model prediction and the observational constraints. We emphasise that the model shown has not been optimised with respect to any likelihood derived from the data. 
We conclude this section by highlighting that while we did not explore in detail the consequences of the LPDE construction on the $S_8$ and Hubble tensions (for a review, see, e.g.,~\cite{Abdalla:2022yfr}), the local nature of the mechanism presented in this work offers an intriguing alternative to the usual solutions for these problems. For instance, the behaviour shown in Fig.~\ref{fig:cosmo_summary} indicates a lower $h(z)$ value at high redshifts and a larger value at low redshifts, which could help alleviate the Hubble tension. 

\section{Discussion and conclusions}
\label{sec:discussion}

In this work, we have proposed the Locally Pumped Dark Energy (LPDE) mechanism, in which late-time cosmic acceleration is triggered by the emergence of non-linear DM structures. The central idea is that short-wavelength, virialised DM modes act as a pump for a scalar field $\chi$, shifting the minimum of its effective potential once a significant fraction of matter has collapsed into halos. At the microphysical level, the activation of the pump is local: it originates from coarse-grained short-mode correlators of the DM field inside virialised objects. However, because the DE field $\chi$ is heavier than the Hubble scale, spatial gradients are exponentially suppressed beyond its Compton wavelength. The field rapidly relaxes to the local minimum of its effective potential, and its stress–energy tensor is vacuum-like. After volume-averaging over the halo population, the cosmological effect is therefore captured by a homogeneous, time-dependent energy density $\bar\rho_\chi(z)$ controlled by the halo volume-filling factor.

The mechanism was originally motivated as a solution to the coincidence problem. In the LPDE scenario, the onset of cosmic acceleration is not tied to the homogeneous matter density $\rho_m(z)$, but rather, to the non-linear collapsed fraction $f_{\rm coll}(z)$. Since the growth of non-linear structure peaks around $z\sim\mathcal{O}(1)$, the activation of DE naturally coincides with the epoch at which galaxies and clusters become abundant. In this sense, the ``why now?'' question is reformulated as a consequence of gravitational collapse rather than as an unexplained tuning of a fundamental cosmological constant.

Although the underlying field $\chi$ is canonical and never violates the null energy condition, the coarse-grained energy density $\bar\rho_\chi(z)$ evolves non-trivially during the activation epoch. As the halo volume-filling fraction grows and the interior pump amplitude increases, $\bar\rho_\chi(z)$ rises rapidly over a finite redshift interval. When interpreted through the conventional continuity equation, this transient growth can lead to an effective EoS parameter that temporarily satisfies $w_{\rm eff}<-1$. Importantly, this apparent phantom behaviour is purely effective: it reflects the time dependence of a vacuum contribution sourced by non-linear structure formation, not a ghost instability or a fundamental negative-kinetic term. The LPDE mechanism, therefore, provides a theoretically controlled example of how an evolving, phantom-like EoS can emerge from a perfectly stable, canonical scalar sector. This feature is particularly intriguing in light of recent low-redshift observations suggesting mild deviations from a pure cosmological constant, including hints of evolving or crossing behaviour in the DE EoS. In the LPDE framework, such behaviour arises naturally from the activation of the pump, without introducing ad hoc time dependence in $w(z)$.

Despite its origin in non-linear structures, the LPDE scenario preserves the large-scale homogeneity and isotropy of the Universe. Because $m_\chi^2 \gg H^2$ whenever the pump is active, fluctuations in $\chi$ are suppressed on cosmological scales by factors of order $k^2/m_\chi^2$. The field tracks its local equilibrium value with negligible spatial variation beyond halo scales, and does not cluster on large scales. Consequently, the background expansion is well described by the standard FLRW framework with a modified, time-dependent DE density. The gravitational dynamics remain those of General Relativity; the influence of non-linear structure enters only through a well-defined EFT for the dark sector.

At the background level, the primary signature of LPDE is a modified late-time expansion history. The model predicts:
\begin{enumerate}
\item a late-time activation of DE correlated with halo formation;
\item a transient deviation of $w_{\rm eff}(z)$ from $-1$ around $z\sim\mathcal{O}(1)$;
\item no early DE and no significant DE clustering.
\end{enumerate}

Because $\bar\rho_\chi(z)$ vanishes at early times, the model automatically satisfies stringent constraints from the CMB, BBN, and high-redshift BAO. The late-time growth of $\bar\rho_\chi$ modifies $H(z)$ at $z\lesssim 1$, and can therefore be tested using supernovae, BAO, cosmic chronometers, and weak-lensing data. A particularly sensitive probe is the late-time integrated Sachs–Wolfe (ISW) effect. The rapid evolution of $\bar\rho_\chi(z)$ during activation enhances the decay of gravitational potentials on large scales, potentially producing a distinctive ISW signature in cross-correlations between CMB temperature maps and large-scale-structure surveys. While current ISW measurements are noise-dominated, future wide-area surveys may provide meaningful constraints. The altered expansion history at $z\lesssim 1$ in the LPDE model influences late-time determinations of $H_0$, potentially shifting inferred values when data are analysed assuming a $\Lambda$CDM template. In addition, since LPDE is tied to non-linear DM structures, the effective DE density grows in proportion to the collapsed fraction of matter. This produces a late-time suppression of structure growth relative to $\Lambda$CDM, providing a natural mechanism to reduce the amplitude of matter clustering and potentially alleviate the $S_8$ tension~\cite{Abdalla:2022yfr}.

A notable strength of the LPDE scenario is its conceptual simplicity. The mechanism follows directly from standard EFT reasoning: once non-linear short modes are integrated out, a tadpole term for a scalar field is generically induced. The resulting displaced minimum produces a vacuum contribution whose magnitude is controlled by a coarse-grained composite operator. In this work, we have adopted an axion-like scalar description of DM as a convenient and well-motivated realisation, since axions naturally provide a scalar field with non-linear clustering. However, the mechanism does not rely on uniquely axionic properties. Any DM sector that develops significant non-linear short-scale fluctuations capable of sourcing a local composite operator can, in principle, generate a similar pumping effect.

The present work is deliberately phenomenological. We have modelled the pump amplitude using a simple parametrisation tied to the halo mass function, and have not yet explored the full range of observational consequences. Thus, several avenues merit further investigation for validating or falsifying the model:
\begin{itemize}
\item A full parameter inference using current cosmological data sets (CMB, BAO, supernovae, weak lensing, and large-scale structure) to determine the viable region of LPDE parameter space.
\item Dedicated studies of the ISW signal and cross-correlations with galaxy surveys.
\item Analysis of the backreaction of the local DE on halo on halo formation, stability, dynamics, and internal halo structure.
\item Implementation of the LPDE mechanism in $N$-body simulations to study the non-linear interplay between halo collapse and the activation of the DE sector.
\item Exploration of more complete microphysical completions and the associated EFT matching conditions.
\end{itemize}

In summary, the LPDE mechanism provides a simple and theoretically motivated framework in which cosmic acceleration is dynamically linked to non-linear structure formation. Originally conceived as a solution to the coincidence problem, it also naturally realises a form of dynamical DE with transient, phantom-like effective behaviour. Given its minimal assumptions and clear observational signatures, the model warrants further theoretical and phenomenological investigation.

\acknowledgments

The authors thank Pedro Ferreira, Jo\~ ao Magueijo, Andrew Tolley, Jesús Torrado, and Toby Wiseman for useful discussions. We thank Matteo Braglia, Marco Peloso, and Angelo Ricciardone for the helpful comments on a late-stage version of the manuscript. M.P. acknowledges the hospitality of Imperial College London, which provided office space during some parts of this project. The work of M.P. is supported by the Comunidad de Madrid under the Programa de Atracción de Talento Investigador with number 2024-T1TEC-31343.

\appendix

\section{Long modes Yukawa suppression}
\label{sec:Yukawa}
Consider the solution for an isolated halo of radius $R$. Eq.~\eqref{eq:chi_static} is solved by a Yukawa Green's function. In physical, spherical coordinates (or setting $a\simeq1$ on the scales of interest), one has
\begin{equation}
G(r)=-\frac{1}{4\pi}\frac{e^{-m_{\chi} r}}{r}.
\end{equation}
The solution for $\chi$ is therefore
\begin{equation}
\label{eq:chi_convolution}
\chi(\mathbf r)
=
m_{\chi}^2
\int \textrm{d}^3x'\,G(|\mathbf r-\mathbf r'|)\,
\chi_{\rm eq}(\mathbf r').
\end{equation}
The response of $\chi$ to a given halo is exponentially suppressed beyond the Compton wavelength $\lambda_\chi\equiv m_{\chi}^{-1}$, ensuring locality of the backreaction.

To make this explicit, consider a spherically symmetric halo $\chi_{\rm eq}(r)=\chi_h\,\Theta(R-r)$, where $\chi_h$ is constant inside the halo. The static field equation becomes
\begin{align}
\frac{1}{r^2}\frac{\textrm{d}}{\textrm{d}r}\!\left(r^2\frac{\textrm{d}\chi}{\textrm{d}r}\right)
- m_{\chi}^2\big(\chi-\chi_h\big) &= 0,
\qquad r<R,\\
\frac{1}{r^2}\frac{\textrm{d}}{\textrm{d}r}\!\left(r^2\frac{\textrm{d}\chi}{\textrm{d}r}\right)
- m_{\chi}^2\chi &= 0,
\qquad r>R,
\end{align}
with regularity at the origin and matching at $r=R$. The interior solution approaches the local equilibrium value, $\chi_{\rm in}(r)\simeq \chi_h$, up to boundary-layer corrections of thickness $\sim\lambda_\chi$. Outside the halo, the solution is Yukawa-suppressed,
\begin{equation}
\chi_{\rm out}(r)\propto \frac{e^{-m_{\chi} r}}{r},
\end{equation}
so the influence of any individual halo does not extend beyond a few $\lambda_\chi$.

\section{LPDE halo density profiles}
\label{sec:DE_density_profile}

A potential concern for the phenomenological pumping mechanism is that the induced DE component inside halos could become dynamically
important compared to the DM distribution that generated it. Since halo formation and evolution are well described by standard CDM dynamics, the pumped component must remain gravitationally subdominant over most of the halo volume.

Using Eqs.~\eqref{eq:rhochi} and~\eqref{eq:fV_identity}, together with the
definition $\bar\rho_{m,h}(z)=200\rho_c(z)$, we obtain the ratio of the
volume-averaged density contributions per halo,
\begin{align}
\eta &\equiv
\frac{\bar\rho_{\chi,h}(z)}{\bar\rho_{m,h}(z)}
=
\frac{\int_{V_h} \rho_{\chi,h}(\mathbf r)\,d^3r}
{\int_{V_h} \rho_{{\rm DM},h}(\mathbf r)\,d^3r},
\label{eq:eta}\\
&=
\frac{1}{f_{\rm coll}(z)}
\frac{\Omega_\Lambda(z)}{\Omega_m(z)} ,
\end{align}
which at present gives $\eta\sim5$. At first sight, this appears problematic, since the halo-averaged LPDE density exceeds the DM counterpart. However, this conclusion holds only if the LPDE and DM components share the same spatial concentration, i.e., if $\phi$ and $\chi$ occupy the same volume. If the LPDE halo is significantly more extended than the DM halo, meaning $V_{h,\chi} > V_{h}$, the enclosed LPDE mass grows much more slowly with radius, and its dynamical impact can remain small throughout most of the halo. 

In the main text, we consider a regime in which the LPDE field is heavier than the Hubble scale, but lighter than $\phi$, corresponding to a Compton wavelength overlapping with the characteristic sizes of galactic and cluster halos, see Eqs.~\eqref{eq:LPDE_mass} and~\eqref{eq:LPDE_wavelength}, respectively. As a result, spatial gradients smooth the LPDE over distances comparable to the halo scale, while DM continues to collapse collisionlessly to much smaller radii. Physically, the DM halo evolves through non-linear gravitational collapse and virialisation on the local dynamical time, producing a concentrated density profile with characteristic scale radius $r_s$. By contrast, the $\chi$ field does not undergo collisionless collapse, but instead, relaxes toward a source-dependent minimum determined by the local pumping term. Therefore, the finite response scale $\lambda_\chi$ smooths the LPDE distribution relative to DM. This naturally leads to a broader LPDE profile with a scale radius
\begin{equation}
r_\chi \sim \max(r_s,\lambda_\chi),
\end{equation}
and hence a smaller effective concentration
\begin{equation}
c_\chi = \frac{R_{200}}{r_\chi} \ll c_{200}.
\end{equation}
Since $\lambda_\chi$ lies within the range of halo scales, we generically
expect $r_\chi \gg r_s$ for galactic halos, leading naturally to
$c_\chi/c_{200}\ll1$.

To obtain an estimate of the required ratio $c_\chi/c_{200}$, we consider the case where halos are described by universal profiles using scale radii. We define the dimensionless radius
$x = r/r_s$, where $r_s$ is the scale radius of the DM halo.
The halo radius is defined by $R_h \equiv R_{200} = c_{200} r_s$. For practical purposes we adopt an Einasto profile~\cite{Einasto1965,Graham:2005xx,2012A&A...540A..70R}
for the DM density,
\begin{equation}
\rho_{\rm E}(x)
=
\rho_s
\exp\!\left[-\frac{2}{\alpha}(x^\alpha-1)\right],
\end{equation}
which provides an accurate description of CDM halos whilst avoiding divergence in the halo core. The mean halo density is
\begin{equation}
\bar\rho_{m,h}
=
\frac{3}{c_{200}^3}
\int_0^{c_{200}} \rho_{\rm E}(x) x^2 dx ,
\end{equation}
and the normalisation of the profile can be fixed by imposing $\bar{\rho}_{m,h}=200\rho_c(z)$
\begin{equation}
\rho_s =
\frac{200}{3} c_{200}^3 e^{-2/\alpha}\rho_c(z)
F_{\rm E}^{-1}(c_{200}),
\end{equation}
where we have introduced the cumulative function
\begin{equation}\label{eq:F_cumul}
F_{\rm E}(x)=\int_0^x
\exp\!\left[-\frac{2}{\alpha}u^\alpha\right]u^2du .
\end{equation}

For the LPDE component, we assume a separate density shape profile
\begin{equation}
\rho_\chi(x)=\mathcal{N}\,\psi(x), 
\end{equation}
giving a mean LPDE density inside $r_\chi$ 
\begin{equation}
\bar\rho_{\chi,h}=\frac{3\mathcal{N}}{c_\chi^3}\,F_\chi(c_\chi),
\end{equation}
where the cumulative function $F_\chi(x)$, for profile $\psi(x)$, is defined as in Eq.~\eqref{eq:F_cumul}.
Using Eq.~\eqref{eq:eta}, the normalisation constant becomes
\begin{align}
\mathcal{N}
&=
\frac{c_\chi^3}{3\,F_\chi(c_\chi)}\eta\bar\rho_{m,h} = \frac{200}{3\,F_\chi(c_\chi)}c_\chi^3\eta\rho_c(z).
\end{align}
Let us stress that while $\bar\rho_{m,h}$ gives the mean matter density of the halo, $\bar\rho_{\chi,h}$ represents the mean LPDE density inside and outside the halo up to $r_{\chi} > R_{200}$. 

The relevant quantity to assess the impact of the LPDE component on the halo dynamics is not the local density ratio $\rho_\chi/\rho_{\rm DM}$, but rather, the relative gravitational acceleration generated by the two components. For a spherically symmetric system, the ratio of radial accelerations at radius $r$ is given by the ratio of the
enclosed masses,
\begin{equation}
\label{eq:ratio_masses}
\mathcal{G}(r)\equiv
\frac{g_\chi(r)}{g_{\rm DM}(r)}
=
2\left|\frac{M_\chi(<r)}{M_{\rm DM}(<r)} \right|,
\end{equation}
where the factor two arises because the gravitational source of a vacuum-like component is proportional to $\rho+3p=-2\rho$. 

We can then express the enclosed LPDE mass as
\begin{equation}
M_\chi(<x)=4\pi r_s^3 \mathcal{N} F_\chi(x),
\end{equation}
and for the Einasto profile, the enclosed DM mass becomes
\begin{equation}
M_{\rm DM}(<x)
=
4\pi r_s^3 \rho_s e^{2/\alpha}F_{\rm E}(x).
\end{equation}
Substituting into Eq.~\eqref{eq:ratio_masses} gives \begin{equation}
\mathcal{G}(x)
=
2\eta\,
\left(\frac{c_\chi}{c_{200}}\right)^3
\frac{F_{\rm E}(c_{200})}{F_\chi(c_\chi)}
\frac{F_\chi(x)}{F_{\rm E}(x)} ,
\end{equation}
where all dimensionful quantities cancel, and the result depends only on the dimensionless radius $x$, the ratio of halo concentrations $c_{200}$ and $c_\chi$, and the density ratio $\eta$.

A useful estimate is obtained near the halo boundary, $x=c_{200}$, where
$F_\chi(x)$ has typically saturated to an $\mathcal{O}(1)$ value while
$F_{\rm E}(x)\sim F_{\rm E}(c_{200})$. In that regime, one finds the simple
scaling
\begin{equation}
\mathcal{G}(R_{200})
\sim
2\,\eta\left(\frac{c_\chi}{c_{200}}\right)^3 .
\label{eq:mu_scaling}
\end{equation}
This makes the parametric suppression particularly transparent: even if the volume-averaged density ratio $\eta$ is of order unity or larger, the dynamical impact of the LPDE component is strongly reduced when its profile is significantly broader than that of the DM.

Assuming that the LPDE and DM components are truncated at the same halo radius $R_{200}$, the difference between the profiles is determined by their scale radii
\begin{equation}
\frac{c_\chi}{c_{200}}=\frac{r_s}{r_\chi}.
\end{equation}
The estimate in Eq.~\eqref{eq:mu_scaling} shows immediately that the LPDE contribution remains dynamically subdominant provided $c_\chi \ll c_{200}$, which corresponds to a broader LPDE profile with $r_\chi\gg r_s$. In this case, the enclosed LPDE mass grows much more slowly with radius, and its gravitational influence remains small throughout most of the virialised region.

To estimate the spatial structure of the LPDE component within halos, we adopt a phenomenological ansatz for the local profile of the pumped vacuum energy. In the EFT description introduced in
Sec.~\ref{subsec:classical_eft_shifted_quad}, the LPDE density satisfies $\rho_\chi \propto \mathcal{J}_h^2$, where $\mathcal{J}_h$ denotes a coarse-grained composite operator constructed from short-distance modes of the DM field. A first-principles calculation of this operator in the non-linear halo environment is beyond the scope of the present work. Instead, we use simple local proxies constructed from the DM field to estimate the relative concentration of the LPDE component with respect to the halo.

In the non-relativistic regime, the energy density of the axion field is
proportional to the square of its oscillation amplitude, implying $\phi_S^2(\mathbf r)\propto \rho_{\rm DM}(\mathbf r)$. This motivates two simple phenomenological models for the LPDE density profile.
\begin{itemize}
    \item If the composite operator is dominated by powers of the field amplitude,
$\mathcal{J}^2_h\propto \phi_S^4$, the LPDE density scales as
\begin{equation}
\label{eq:rho_phi2}
\psi(x)\propto \rho_{\rm E}^2(x).
\end{equation}
\item Alternatively, if the pumping is dominated by the gradient energy of the axion field, one may take $\mathcal{J}_h^2\propto(\nabla\phi_S)^{4}$. Using $\phi_S^2(\mathbf r)\propto \rho_{\rm DM}(\mathbf r)$, this gives
\begin{equation}
\rho_\chi(\mathbf r)
\propto
\frac{[\nabla\rho (\mathbf r)]^{2 }}{\rho^{2 } (\mathbf r) }.
\end{equation}
For an Einasto halo, this leads approximately to
\begin{equation}\label{eq:J2_psi}
\psi(x)\propto \rho_{\rm E}^2(x)\,x^{4(\alpha-1)}.
\end{equation}
\end{itemize}
Both choices provide illustrative examples of how the LPDE density may be distributed within halos depending on the dominant microscopic operator.

For typical halo values $\alpha\simeq0.15$--$0.25$, the acceleration ratio $\mathcal{G}(r)$ increases toward the centre, but remains small over most of the halo volume when $c_\chi/c_{200}\ll1$. In this regime, the pumped component may dominate the density only within a very small inner region, while remaining dynamically subdominant throughout the bulk of the halo. 

\begin{figure*}[ht!]
  \centering
\includegraphics[width=\textwidth]{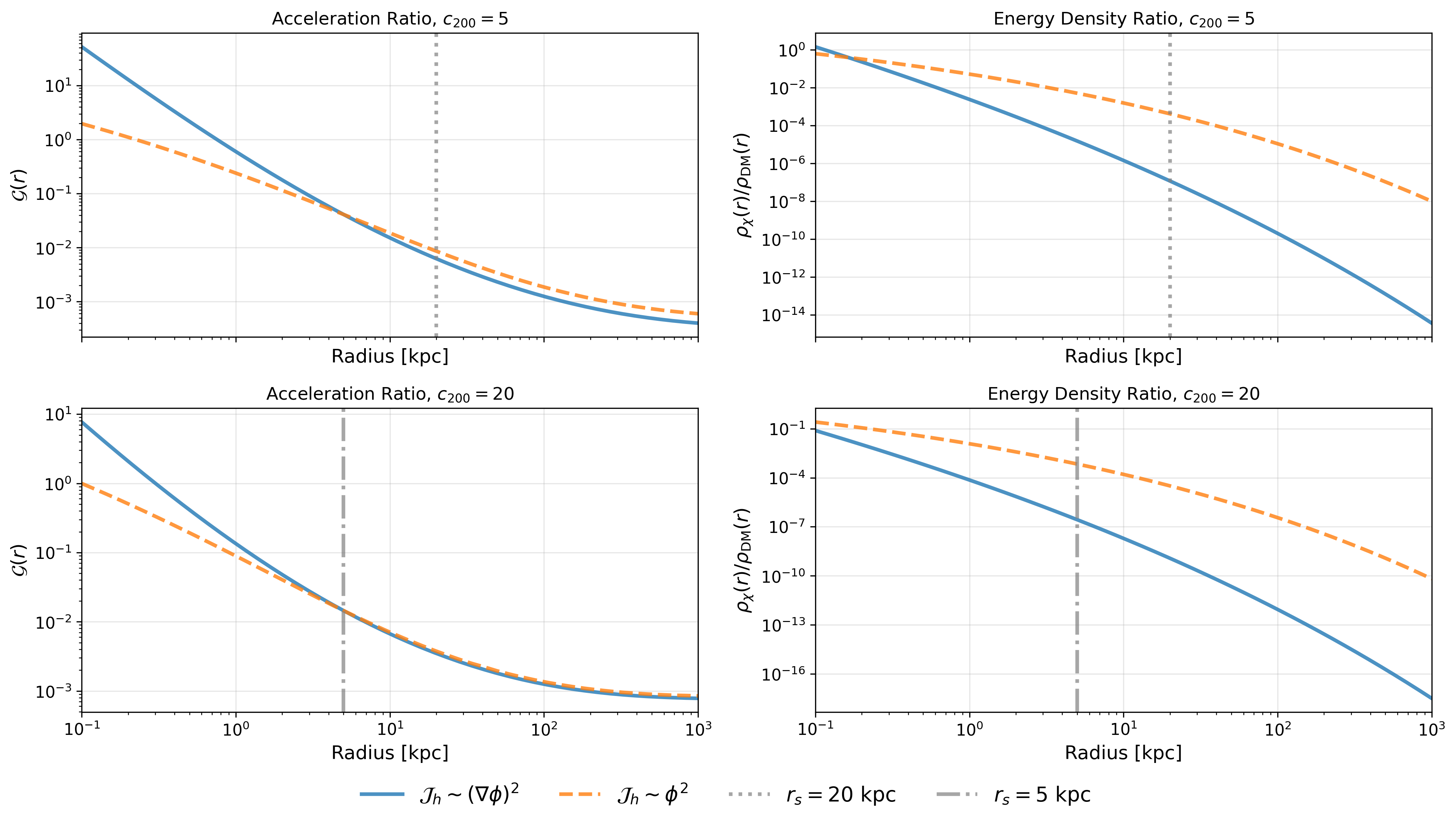}
  \caption{Acceleration ratio $\mathcal{G}(r)$ (left column) and density ratio $\rho_\chi/\rho_{\rm DM}$ (right column) for two representative galactic halo with Einasto profiles parameters $\alpha=0.15$, $r_s=20\,{\rm kpc}$, $c_{200}=5$, (top row) and $\alpha=0.15$, $r_s=4\,{\rm kpc}$, $c_{200}=20$, (bottom row), both corresponding to $c_{200}/c_\chi=20$. Both the amplitude-dominated (solid blue lines) and the gradient-dominated (dashed orange) profiles are shown.}
  \label{fig:mass_density_ratio}
\end{figure*}

In Fig.~\ref{fig:mass_density_ratio}, we show the mass (left panels) and energy-density (right panels) ratios for a galaxy-like halo with parameters $r_s=20\,{\rm kpc}$, $\alpha=0.15$ and $c_{200}=5$ (top row), and $r_s=4\,{\rm kpc}$, $\alpha=0.15$ and $c_{200}=20$ (bottom row), both corresponding to $c_{200}/c_\chi=20$, with LPDE profile constructed from either choice, Eqs.~\eqref{eq:J2_psi} or~\eqref{eq:rho_phi2}. The limiting density ratio is fixed to $\eta=5$. For these parameters, the LPDE component contributes only a small fraction of the gravitational acceleration over most of the halo volume. In this example, the acceleration ratio becomes of order unity only within the inner core, around $r\sim1\,{\rm kpc}$. This behaviour illustrates how a sufficiently extended LPDE profile can remain dynamically subdominant throughout the bulk of the halo despite the large volume-averaged density ratio $\eta$.

At the same time, the enhanced contribution in the central region suggests that the LPDE component could modify the dynamics of the inner halo. While our simple static example is not intended as a full dynamical model, it highlights how the LPDE mechanism might lead to observable signatures in the central structure of halos. In particular, the presence of a smoothed vacuum-like component could potentially influence the formation of central density cusps or the internal dynamics of dwarf galaxies, issues that are known to present challenges for standard CDM structure formation (see, e.g.,~\cite{Bullock:2017xww,2019ARA&A..57..375S}).

For the parameters explored here, the LPDE contribution modifies the halo gravitational field at the virial radius by $\mathcal{G}(R_{200})\sim10^{-3}$. This level of suppression is far below current weak-lensing and galaxy–galaxy lensing sensitivities, implying that the LPDE component would primarily affect only the innermost halo dynamics~\cite{Kilbinger:2014cea,Mandelbaum:2017jpr}. Alternatively, if the DE profiles develop substantial extended tails, with $\mathcal{G}(R_{200}) \sim \mathcal{O}(10^{-1})$, large-scale structure measurements would infer an effective matter density biased high relative to CMB constraints. This reflects the contribution of clustered DE to the gravitational potential without a corresponding increase in the matter density, leading to a reduced inferred clustering amplitude and potentially alleviating the $S_8$ tension.

Our argument does not attempt to model the full coupled evolution of the DM and LPDE components. Therefore, it should only be regarded as an illustrative estimate. Nevertheless, it provides a useful guide to the mass scale required for the LPDE field. In the example above, the ratio $c_{200}/c_\chi=20$ corresponds approximately to an LPDE halo scale $R_h^\chi\sim400\,{\rm kpc}$.

A further consistency condition is that the gradient energy of the $\chi$
configuration remains subdominant to the vacuum-like contribution inside the halo. For a halo of radius $R^\chi_h$, the Yukawa solution implies that the field
interpolates to the exterior over a shell of thickness
$\lambda_\chi=m_\chi^{-1}$, so that the ratio of total gradient to interior
vacuum energy scales as $\mathcal O(\lambda_\chi/R^\chi_h)$.
Hence, the volume-averaged treatment is self-consistent provided
$r_s < \lambda_\chi\ll R^\chi_h$. For the example above, this gives an estimate of $m_\chi\sim 10^6 \,H_0$ which lies within the intermediate mass range discussed in section~\ref{subsec:chi_green_halos}.

\section{Parameter dependence}
\label{sec:parameter_dependence}

\begin{figure*}[ht!]
  \centering
  \includegraphics[width=\textwidth]{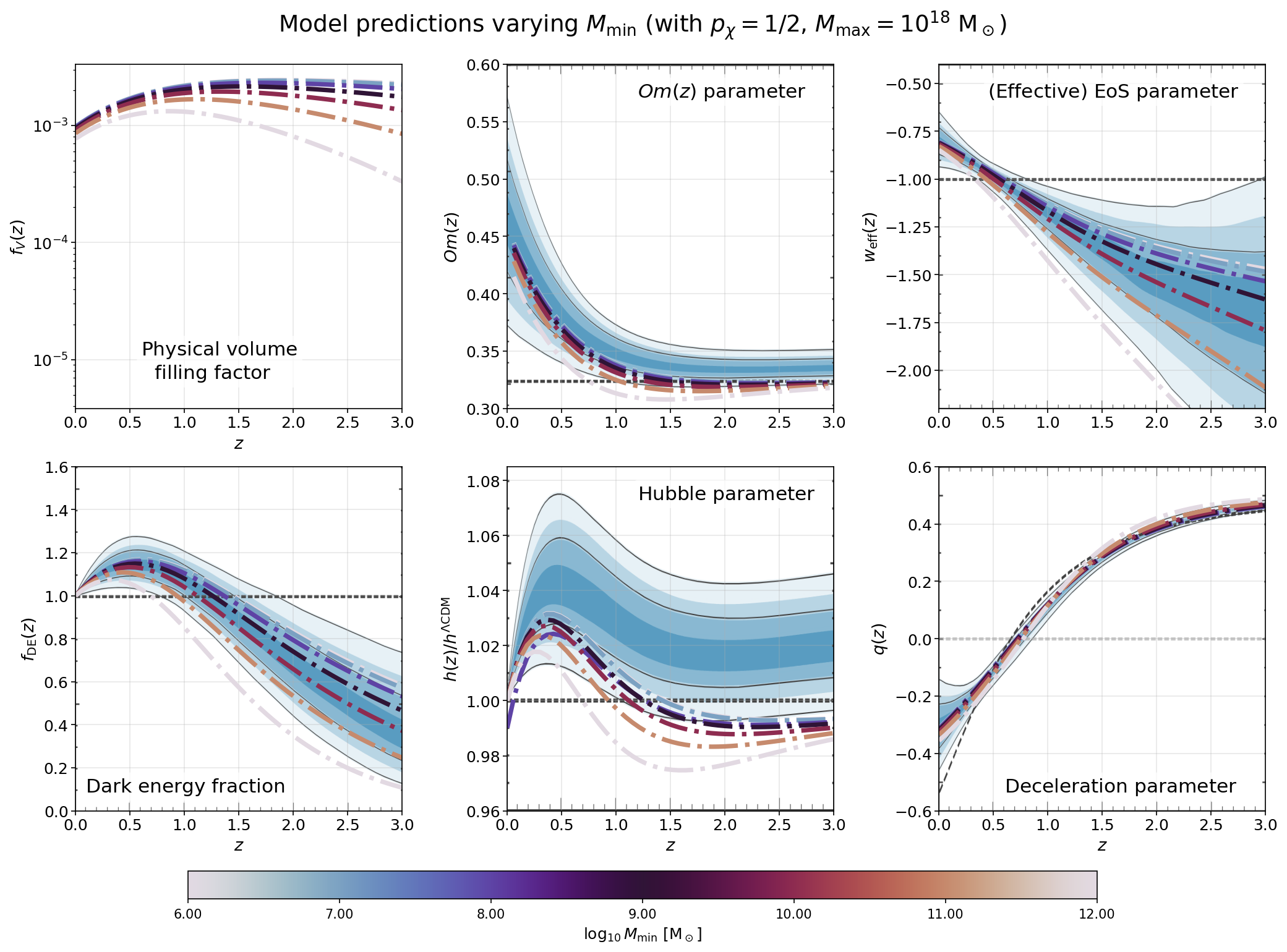}
  \caption{
  Same plots as in Fig.~\ref{fig:cosmo_summary}, but varying $M_{\rm min}$ in the range $[10^{5} M_{\odot}, 10^{12} M_{\odot}]$. The other parameters are fixed to $p_{\chi} = 1/2$ and $M_{\rm max} = 10^{18} M_{\odot}$.}
  \label{fig:cosmo_vary_Mmin}
\end{figure*}

In this appendix, we discuss the impact of varying the model parameters on the observable predictions shown in Fig.~\ref{fig:cosmo_summary}.

In Fig.~\ref{fig:cosmo_vary_Mmin}, we show the impact of varying $M_{\rm min}$, in particular, we choose values in the range $M_{\rm min} \in [10^{5} M_{\odot}, 10^{12} M_{\odot}]$, while keeping fixed the other two parameters fixed to $p_{\chi} = 1/2$ and $M_{\rm max} = 10^{18} M_{\odot}$. We notice that for smaller values of $M_{\rm min}$, the activation of the DE component starts earlier, and the transition is smoother. This is because smaller halos start to contribute to the collapsed fraction at higher redshifts, leading to an earlier and more gradual growth of the DE density. Conversely, larger values of $M_{\rm min}$ delay the onset of DE activation and produce a sharper transition, as only more massive halos contribute to the collapsed fraction at later times. A smaller (larger) value of $M_{\rm min}$ corresponds to a shallower (steeper) growth of $w_{\rm eff}$ with time (i.e., for decreasing $z$), this corresponds to a larger (smaller) DE density at earlier times, and therefore, to a smaller (larger) deviation from $\Lambda$CDM. Finally, we notice that $q(z)$ and $Om(z)$ are only marginally affected by the variation of $M_{\rm min}$, while the other quantities are more sensitive to this parameter.

\begin{figure*}[t]
  \centering
  \includegraphics[width=\textwidth]{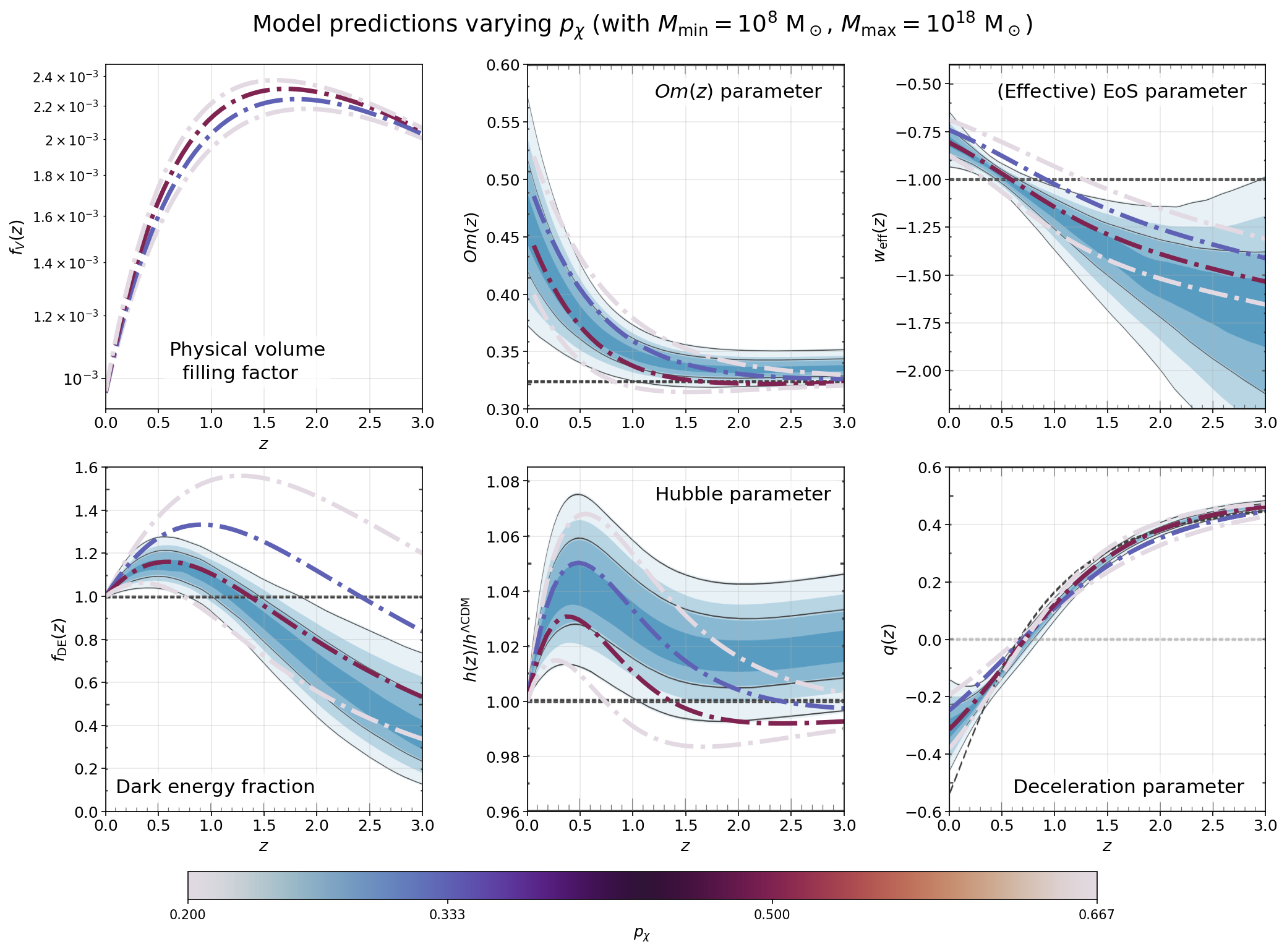}
  \caption{
  Same plots as in Fig.~\ref{fig:cosmo_summary}, but varying $p_{\chi}$ in the range $[1/5, 2/3]$. The other parameters are fixed to $M_{\rm min} = 10^{8} M_{\odot}$ and $M_{\rm max} = 10^{18} M_{\odot}$.}
  \label{fig:cosmo_vary_pchi}
\end{figure*}

Let us move to Fig.~\ref{fig:cosmo_vary_pchi}, where we show the impact of varying $p_{\chi}$. In particular, we choose values in the range $p_{\chi} \in [1 / 5, 2/3]$, while fixing $M_{\rm min} = 10^{8} M_{\odot}$ and $M_{\rm max} = 10^{18} M_{\odot}$. In this case, a smaller (larger) value of $p_{\chi}$ leads to a slower (faster) growth of the DE density. In the case of a small $p_{\chi}$, the pump is less efficient, and therefore, to get the correct DE density, the pump needs to be active for a longer time, implying that the $\chi$ field starts to contribute to the energy density at earlier times. This behaviour is quite evident from the bottom-left panel showing the DE fraction. Concerning the evolution of $w_{\rm eff}$, the smallest value of $p_{\chi}$ shown in this plot produces a behavior that closely resembles a $w_0w_a$CDM cosmology, while for larger values, we clearly see a break from this simple parametrisation at low redshifts, with a more rapid growth of $w_{\rm eff}$. Compared to Fig.~\ref{fig:cosmo_vary_Mmin}, we notice that the variation of $p_{\chi}$ impacts all parameters, with $q(z)$ remaining the least affected. 

As a final comment, since we have noticed that the variation of $M_{\rm max}$ has a negligible impact on the results, we do not show it here.

\newpage
\bibliography{pumped_de}

\end{document}